\documentclass[pra,twocolumn,eqsecnum,]{revtex4}
\usepackage{graphicx}
\usepackage[urlcolor=blue, hyperindex, colorlinks, bookmarks=true,linkcolor=black,citecolor=black]{hyperref}

\begin{document}


\newcommand{\sch}{Schr\"odinger }
\newcommand{\schs}{Schr\"odinger's }
\newcommand{\nn}{\nonumber}
\newcommand{\nl}{\nn \\ &&}
\newcommand{\dg}{^\dagger}
\newcommand{\bra}[1]{\langle{#1}|}
\newcommand{\ket}[1]{|{#1}\rangle}
\newcommand{\Bra}[1]{\Big{\langle}{#1}\Big{|}}
\newcommand{\Ket}[1]{\Big{|}{#1}\Big{\rangle}}
\newcommand{\bl}{{\Bigl(}}
\newcommand{\br}{{\Bigr)}}
\newcommand{\erf}[1]{Eq.~(\ref{#1})}
\newcommand{\erfs}[2]{Eqs.~(\ref{#1}) and (\ref{#2})}
\newcommand{\erft}[2]{Eqs.~(\ref{#1}) -- (\ref{#2})}
\newcommand{\beq}{\begin{equation}}
\newcommand{\eeq}{\end{equation}}


\title{Modal dynamics for positive operator measures}
\author{Jay Gambetta}
\affiliation{Centre for Quantum Dynamics, School of Science,
Griffith University, Brisbane 4111, Australia}
\author{H. M. Wiseman} \email{h.wiseman@griffith.edu.au}
\affiliation{Centre for Quantum Dynamics, School of Science,
Griffith University, Brisbane 4111, Australia} 

\begin{abstract}
The modal interpretation of quantum mechanics allows one to keep
the standard classical definition of realism intact. That is,
variables have a definite status for all time and a measurement
only tells us which value it had. However, at present modal
dynamics are only applicable to situations that are described in
the orthodox theory by projective measures. In this paper we
extend modal dynamics to include positive operator measures
(POMs). That is, for example, rather than using a complete set of
orthogonal projectors, we can use an overcomplete set of
nonorthogonal projectors. We derive the conditions under which
Bell's stochastic modal dynamics for projective measures reduce to
deterministic dynamics, showing (incidentally) that Brown and
Hiley's generalization of Bohmian mechanics [quant-ph/0005026,
(2000)] cannot be thus derived. We then show how {\em
deterministic} dynamics for positive operators can also be
derived. As a simple case, we consider a Harmonic oscillator, and
the overcomplete set of coherent state projectors (i.e. the Husimi
POM). We show that the modal dynamics for this POM in the
classical limit correspond to the classical dynamics, even for the
nonclassical number state $\ket{n}$. This is in contrast to the
Bohmian dynamics, which for energy eigenstates, the dynamics are
always non-classical.
.\\\\
\end{abstract}

\maketitle

\section{INTRODUCTION}

It has been almost eighty years since quantum theory emerged as a
complete theory, yet fundamental debates still occur over its
interpretation. These debates usually center around what has been
called the measurement problem, which we argue is a twofold
problem.

In quantum mechanics a complex vector $\ket{\Psi(t)}$, which
belongs to a Hilbert space ${\cal H}$, is used to describe the
state of a system, and its evolution is given by the \sch
equation. In the orthodox interpretation it is postulated that a
system only has a definite value for an observable $\hat{A}$
(energy, position etc.) if the state $\ket{\Psi(t)}$ is an
eigenstate, $\ket{\Psi_{n}(t)}$, of $\hat{A}$, in which case the
value of $\hat{A}$ is the eigenvalue $a_{n}$ associated with
$\ket{\Psi_{n}(t)}$. However, since the \sch equations in linear
it is possible for the system to be in a state which is a linear
superposition of these eigenstates:
$\ket{\Psi(t)}=\sum_{n}c_{n}(t)\ket{\Psi_{n}(t)}$. That is, the
system can have two or more values of an observable at once or, we
say, the value is not defined. To explain why only one of the
values is obtained in a measurement we need to introduced an extra
dynamical equation, the reduction equation. This equation for a
measurement of observable $\hat{A}$ has the effect of collapsing
the state, $\ket{\Psi(t)}\rightarrow\ket{\Psi_{n}(t)}$, thereby
defining the value of the observable. Thus the standard view of
reality (that variables have well defined values even when they
are not observed) is lost. This is what we call the first problem
of the measurement problem under the orthodox interpretation.

The second problem is concerned with choices: what observable is
measured, and where the reduction occurs (the Heisenberg cut
{\cite{Hei}). In the orthodox interpretation there is a classical
world, which we live in, existing outside of the quantum world and
allows us to define an apparatus which chooses the observable to
be measured (the set of eigenstates to be collapsed into). The
problem is, at what point do we place the distinction between
system and apparatus? An apparatus is made of particles just like
the system. Perhaps it is only an intelligent observer that
collapses the wavefunction. But it is not clear how intelligence
or consciousness can influence physics. Alternatively it can be
argued that quantum mechanics should extend up the von Neumann
chain \cite{Von} to include the entire universe, so
$\ket{\Psi(t)}$ now labels the state of the universe. If this is
the case then what is external to the universe which makes the
quantum measurements?

In this paper we consider the modal interpretation of quantum
mechanics
\cite{Bel84,Sud87,Vin93,Bub97,Sud00,Dic97,BacDic99,VerDie95,Die95,Die97,
Hea89,SpeSip01,SpeSip01b}. This interpretation's central goal is
to keep the standard definition of reality intact, that
observables have values even when not observed. In this
interpretation to explain why only one value of $\hat{A}$ is
actual for all time we introduce an extra state, the property
state, $\ket{{\Psi}_{n}(t)}$. This state evolves stochastically
with time (jumping between different values of $n$) and selects
the present value of $\hat{A}$, $a_n$, from the set of possible
values $\{a_n\}$. While $\ket{\Psi(t)}$ in effect chooses the
weights for this stochastic jumping; that is, it acts as a guiding
state. We define the unnormalized (signified by a tilde) property
state as
\begin{equation}\label{unnormpropertyState}
  \ket{\tilde{\Psi}_{n}(t)}={\hat{\pi}_{n}(t)\ket{\Psi(t)}},
\end{equation} where $\hat{\pi}_{n}(t)$ is a projector (acting in the total Hilbert space
of the universe ${\cal H}_{\rm uni}$) and satisfies
$\sum_{n}\hat\pi_{n}(t)=\hat{1}$. The set of projectors,
$\{\hat{\pi}_{n}(t)\}$ are labelled the preferred projective
measure. The normalized property state is then defined as
\begin{equation}\label{propertyState}
  \ket{{\Psi}_{n}(t)}={\hat{\pi}_{n}(t)\ket{\Psi(t)}} /{\sqrt{N}},
\end{equation}
where ${N}$ is a normalization constant. Note the set of property
states depends on both the preferred projective measure and the
guiding state. The guiding state $\ket{\Psi(t)}$ (as in the
orthodox interpretation) is found from the \sch equation
\begin{equation}\label{sch}
  d_{t}\ket{\Psi(t)}=-i\hat{H}_{\rm uni}(t)\ket{\Psi(t)},
\end{equation} where $\hat{H}_{\rm uni}(t)$ is the Hamiltonian of
the universe.

The reason why only one value for $\hat{A}$ is obtained upon a
measurement of $\hat{A}$ at time $t$, is we choose the property
states (i.e. choose $\hat{\pi}_{n}(t)$) so that $\hat{A}$ can be
written as $\sum_{n}a_n\hat{\pi}_n(t)$. Now if $\ket{\Psi_{n}(t)}$
is the property state at time $t$, then $\hat{A}$ has the value
$a_n$ at $t$ and if $\hat{A}$ is measured, this value is revealed.
The probability that universe will be in the $n^{\rm th}$ property
state at time $t$ is given by
\begin{equation}\label{Prob}
  P_{n}(t)=\bra{\Psi(t)}\hat\pi_{n}(t)\ket{\Psi(t)}=\langle{\tilde{\Psi}_{n}(t)}\ket{\tilde{\Psi}_{n}(t)},
\end{equation} the standard Born probability. Thus the average of $\hat{A}$ over all possible
property states will agree with $\langle\hat{A}\rangle$ found by
the orthodox theory. However, unlike the orthodox interpretation
this probability does not refer to the probability of observing
results $a_n$ at time $t$ but to the probability of $A$ having
value $a_n$ at time $t$.

For a given set of property states, there will be more than one
observable, which can be assigned a definite value. Such and
observable should, and from now on will, be referred to as a
property or beable (after Bell \cite{Bel84}). Our notation for a
property is
\begin{equation} \label{projproperty}
  {Z}=\{(z_{n}, \hat{\pi}_{n}(t))\}
\end{equation}
where $z_{n}$ is the value of the property, which could be a real
number, or a complex number, a statement (yes/no) or even a string
of numbers. When the set $\{z_n\}$ are real (complex) numbers the
property can be represented by a Hermitian (normal) operator
\begin{equation}\hat{Z}(t)=\sum_{n}z_n\hat{\pi}_n(t).
\end{equation}

Thus, in this interpretation, the first problem of the measurement
problem is solved, at least for measurements of the preferred
observable (property), as a measurement only tells us which value
was possessed. For measurements of other observables see the
discussion by Bohm for their interpretation \cite{Boh}. However,
the problem of choice still remains as we can choose a different
preferred projective measure, $\{\hat\pi_{n}(t)\}$, which will
give a different group of observables property status. This
problem for modal interpretations has not been resolved in a
wholly satisfactory way, and may never be.

The problem of choice is illustrated by the many variants of the
modal interpretation. In the beable variant
\cite{Bel84,Sud87,Vin93,Bub97,Sud00} choice of the preferred
projective measure is fundamental. That is, it is made
independently of the guiding state. The problem is that many
choices are viable \cite{Bub97}. In other modal theories
\cite{Dic97,BacDic99,VerDie95,Die95,Die97} the preferred
projective measure depends on $\ket{\Psi(t)}$. These have been
labelled by Bacciagaluppi and Dickson \cite{BacDic99} as the
atomic version of the modal interpretation. One assumes a
preferred factorization of the universe ${\cal
H}_{\alpha}\otimes{\cal H}_{\beta}\otimes...\otimes{\cal
H}_{\omega}$. (This seems to be necessary to avoid a
Kochen-Specker type contradiction \cite{Bac95}). In this preferred
factorization the spectral resolution of each subsystems
$\rho^{\alpha}_{\rm red}(t)=\sum_{n_i}w_{n_i}
\hat{\pi}^{\alpha}_{n_i}(t)$ defines the
 preferred projective measure for the universe as
\begin{equation}
  \{\hat{\pi}_{n}(t)=\hat{\pi}^{\alpha}_{n_i}(t)\hat{\pi}^{\beta}_{n_j}
  (t)...\hat{\pi}^{\omega}_{n_z}(t)\},
\end{equation}
where $n=(n_i,n_j,..., n_z)$. This may seem like it has rectified
the problem of choice but it still remains here in the choice of
the preferred factorization.  Another problem of choice can emerge
in the atomic variant this being that for some guiding states
there are many viable algorithms for defining the preferred
projector. Note other variants of the modal interpretation
\cite{Bub97,Hea89,SpeSip01b} which try to answer choice also
encounter the above problems.

In this paper we do not propose to solve the problem of choice. To
the contrary, we show that there is actually {\em more} choices to
be made than what was previously realized. Specifically, it is not
necessary to restrict the property to be of the form of
\erf{projproperty}, based upon orthogonal projectors. The extra
choice is motivated by the fact that in the orthodox
interpretation of quantum mechanics we can extend the theory of
measurement from that using projective measures to that using
positive operator measures (POMs).  The former  measurement (of
observable Z) is described by a set of pairs as in
\erf{projproperty}, but here the $z_n's$ would be interpreted as
measurement results.  The latter measurement (of observable Z) is
described by a set of pairs
 \begin{equation} \label{POMobservables}
  {Z}=\{(z_{n}, \hat{F}_{n}(t))\},
\end{equation}
where $\hat{F}_{n}(t)$ is referred to as a POM element
\cite{Hel76} or effect \cite{Kaus83}. These effects are positive
and complete, with $\sum_{n}\hat{F}_{n}(t)=\hat{1}$. In this
paper, we show that we can develop modal dynamics for preferred
observables of the form in \erf{POMobservables}.

To agree with orthodox theory, our modal dynamics for POMs should
generate the usual probability formula for POMs, which is
\begin{equation}\label{ProbPOM}
  P_{n}(t)=\bra{\Psi(t)}\hat F_{n}(t)\ket{\Psi(t)}.
\end{equation}
Note that in general the number of possible results ($\sum_{n}1$)
can be greater than the dimension of ${\cal H}_{\rm uni}$. Some
examples of POM measurements include informationally complete POMs
\cite{Fuchs,Info} and the Husimi or $Q$-function POM
\cite{Husimi,WalMil94}. In the orthodox theory the reduction
equation for POM-type measurement is
$\ket{\Psi}\rightarrow\hat{M}_{n}\ket{\Psi}$ where $\hat{M}_{n}$
is referred to as the measurement operator and is defined such
that $\hat F_{n}(t)=\hat{M}_{n}\dg\hat{M}_{n}$ \cite{Kaus83}.

Since the modal dynamics presented in Refs.~
\cite{Bel84,Sud87,Vin93,Sud00,Dic97,BacDic99} assume that the
property states are distinguishable, non-orthogonal states are not
allowed in their analysis. In this paper we use Naimark's theorem
\cite{Hel76} to extend the dynamics to include POMs. We begin in
Sec. \ref{Sec2} by extending the stochastic dynamics introduced by
Bell \cite{Bel84}. In Sec. \ref{Sec3} we consider the limits in
which deterministic dynamics can be derived from the stochastic
dynamics, for both the projective and the POM case. Along the way
we show that Brown and Hiley's \cite{BroHil01} generalization of
Bohmian mechanics cannot be derived from Bell's modal dynamics. In
Sec. \ref{Sec4} we illustrate the POM case with an example of a
universe consisting of only a single Harmonic oscillator. We show
that when the Husimi POM (the overcomplete set of coherent state
projectors), is used, the modal dynamics is consistent with
classical mechanics in the classical (large excitation) limit.
However when the modal dynamics is used to describe position
property states (i.e. Bohmian mechanics) there is no recovery of
classical behaviour in the classical limit. This illustrates the
potential usefulness of modal dynamics of POMs.

\section{DYNAMICS FOR MODAL INTERPRETATIONS} \label{Sec2}

In this section we first briefly outline the modal dynamics for
the projective case, which is basically a reproduction of the
results presented in Refs. \cite{Bel84,Sud87,Vin93} and
generalized in Refs. \cite{Sud00,Dic97,BacDic99} to include
time-dependent projectors. Secondly we use Naimark's theorem
\cite{Hel76} to extend the dynamics to include positive operators.

\subsection{The Projective Case} \label{Sec2a}

Before reproducing the standard results for the projector case we
 outline the general method used to describe a stochastic
process, that jumps between $N$ distinct states. We define
$P_{n}(t)$ as the probability that the system is in the $n^{th}$
state at time $t$. Assuming a Markovian process, by which we mean
that the probability of being in state $m$ at time $t+dt$ only
depends on the state at time $t$, we can
write the master equation as
\begin{equation}\label{Master}
  d_t P_{n}(t)=\sum_{m}[T_{nm}(t)P_{m}(t)-T_{mn}P_{n}(t)],
\end{equation} where $T_{nm}$ is defined as
\begin{equation}\label{t}
  T_{nm}(t)=[P_{nm}(t+dt,t)-P_{nm}(t,t)]/dt.
\end{equation} Here $P_{nm}(t+dt,t)$ is a conditional probability and is read as the probability of
the system being in state $n$ at time $t+dt$ given it was in state
$m$ at time $t$. From this definition it follows that $
\sum_{n}T_{nm}=0.$

For distinct states the conditional probability $P_{nm}(t,t)$ must
be 0 for $n\neq m$. This allows us to interpret
 $T_{nm}(t)dt$ for $n\neq m$ as the
transition probability, so we call $T_{nm}(t)$ the transition
rates. For $n=m$, $P_{nn}(t,t)=1$ and $T_{nn}$ (which is negative)
is a measure of the rate at which state $n$ losses probability.

Defining a probability current $J_{nm}(t)$ as
\begin{equation}\label{probcurrent}
J_{nm}(t)=T_{nm}(t)P_{m}(t)-T_{mn}P_{n}(t),
\end{equation} results in $J_{nm}(t)=-J_{mn}(t)$ and
allows us to rewrite the probability master equation as
\begin{equation} \label{master}
d_t P_{n}(t)=\sum_{m}J_{nm}(t).
\end{equation} Given $J_{nm}(t)$ and $P_n(t)$, there
are many possible  transition rates satisfying \erf{master}. One solution,
chosen by Bell \cite{Bel84} is as follows.

For $J_{nm}(t)<0$,
\begin{eqnarray}\label{t1}
  T_{nm}(t)&=&0, \\
  T_{mn}(t)&=&-J_{nm}(t)/P_{n}(t),
\end{eqnarray} and for $J_{nm}(t)>0$
\begin{eqnarray}
  T_{nm}(t)&=&J_{nm}(t)/P_{m}(t),\\
  T_{mn}(t)&=&0. \label{t4}
  \end{eqnarray}
This is only one of the infinitely many solutions. These are found
by adding an extra term, $T^{0}_{nm}(t)$, to $T_{nm}(t)$, where
$T^{0}_{nm}(t)$ is constrained only by
\begin{equation}
T_{nm}^{0}(t)P_{m}(t)-T_{mn}^{0}(t)P_{n}(t)=0.
\end{equation}

To make the link with quantum mechanics we say that the $N$
distinct states are the property states $\{\ket{\Psi_{n}(t)}\}$,
and the set of properties which have definite values form the
group ${\cal G}$, defined by elements
\begin{equation}
 \hat{Z}=\sum_{n}z_n\hat{\pi}_n(t),
\end{equation}
with the value of property $\hat{Z}$ being the corresponding
$z_n$. The evolution of these values (jumping between $z_{n}$) is
determined by the rates $T_{nm}(t)$, which themselves depend upon $J_{nm}(t)$.

By taking the time derivative of \erf{Prob} we obtain the
differential equation
\begin{eqnarray}\label{step1}
d_{t} P_{n}(t)&=&2{\rm Im}[\bra{\Psi(t)}\hat{\pi}_{n}(t)\hat{H}_{\rm
uni
}(t)\ket{\Psi(t)}]\nl+\bra{\Psi(t)}d_{t}[{\hat{\pi}}_{n}(t)]\ket{\Psi(t)},
\end{eqnarray} where we have used \erf{sch}, the \sch equation.
This can be simplified by defining the Hermitian operator,
$\hat{R}(t)$, for which
\begin{equation}
  d_{t}\hat{\pi}_{n}(t)=-i[\hat{R}(t),\hat{\pi}_{n}(t)].
\end{equation} This allows us to rewrite \erf{step1} as
\begin{equation}
d_{t} P_{n}(t)=2{\rm Im}\{\bra{\Psi(t)}\hat\pi_{n}(t)[\hat{H}_{\rm
uni }(t)-\hat{R}(t)]\ket{\Psi(t)}\}.
\end{equation}
Comparing this with \erf{master} and using the fact that
$\sum_{m}\hat\pi_{m}(t)=\hat{1}$, one possible probability current
(that was chosen by Bell \cite{Bel84}) is
\begin{equation}
  J_{nm}(t)=2{\rm Im}\{\bra{\Psi(t)}\hat{\pi}_{n}(t)[\hat{H}_{\rm uni
}(t)-\hat{R}(t)]\hat{\pi}_{m}(t)\ket{\Psi(t)}\}.
\end{equation} Note that this is only one of  infinitely many possible
currents, as we can add any current $J_{nm}^{~0}(t)$ to
$J_{nm}(t)$ which satisfies $  \sum_{m} J_{nm}^{~0}=0,$
 to give a valid probability current. For the purposes of this paper we
only consider the simple solutions (not containing the extra
$T_{nm}^{0}(t)$ and $J_{nm}^{~0}(t)$ terms). For a discussion
about these solution see \cite{Vin93} and \cite{Dic97}.

\subsection{The Positive Operator Measure Case}

A typical example of a POM is the observable
\begin{equation} \label{POMobservablesa}
  {Z}=\{(z_{n},\frac{1}{B}\ket{z_n}\bra{z_n} )\},
\end{equation}
where $\{\ket{z_n}\}$ forms an overcomplete basis in ${\cal
H}_{\rm uni}$, with $\sum_n \ket{z_n}\bra{z_n} = \hat{1}B$. If the
universe was in $\ket{z_n}$, then a measurement of this observable
could with finite probability yield any of the values $\{z_{n}\}$.
This clearly disagrees with the above description of modal
dynamics. Thus we cannot treat non-orthogonal states in the same
manner as orthogonal states.

POMs are not always generated from non-orthogonal states. Given
any set of projectors $\{\hat{\Pi}_{n}(t)\}$ in a larger Hilbert
space ${\cal K}={\cal H}_{\rm uni}\otimes{\cal H}_{\rm aux}$ where
${\cal H}_{\rm aux}$ is some auxiliary Hilbert space, it is well
know that a set of POM elements can always be found by
\cite{Hel76}
\begin{equation}
  \hat{F}_{n}(t)={\rm Tr}_{\rm aux}[(\hat{1}_{\rm uni}\otimes\hat{\rho}_{\rm aux})\hat{\Pi}_{n}(t)]
\end{equation}
where $\hat{\rho}_{\rm aux}$ is a state in ${\cal H}_{\rm aux}$.
For simplicity we define it as $\hat{\rho}_{\rm
aux}=\ket{\phi}\bra{\phi}$. What is perhaps less well known is
that given a POM $\{F_{n}(t)\}$ it is always possible to generate
a projective measure $\{\hat{\Pi}_{n}(t)\}$ in a larger Hilbert
space ${\cal H}_{\rm uni}\otimes{\cal H}_{\rm aux}$, where ${\rm
dim}({\cal H}_{\rm aux})$ is not necessarily equal to $\sum_{n}1$.
This is called Naimark's theorem \cite{Hel76,BusGraLah95}. It
basically says that we can define a projector $\hat{\Pi}_{n}(t)$
such that
\begin{eqnarray} \label{Nai}
&&{\rm Tr}_{{\cal H}_{\rm
uni}}[\ket{\Psi(t)}\bra{\Psi(t)}\hat{F}_{n}(t)]\nl={\rm Tr}_{\cal
K}[\ket{\Psi(t)}\bra{\Psi(t)}\otimes\ket{\phi}\bra{\phi}\hat{\Pi}_{n}(t)],
\end{eqnarray} for all $\ket{\Psi(t)}\in{\cal H}_{\rm uni}$ and
for $n=1,...,N$ where $N$ is the number of POM elements in the
POM. To work out the set $\{\hat{\Pi}_{n}(t)\}$ it is necessary to
introduce another projector $\hat{\Pi}_{N+1}(t)$, such that
\begin{eqnarray}\label{pro1}
\sum_{n}^{N+1}\hat{\Pi}_{n}(t)&=&\hat{1}_{\rm uni+ aux},
\end{eqnarray}
and \begin{eqnarray}
 \label{pro2}
\hat{\Pi}_{n}(t)\hat{\Pi}_{m}(t)&=&\hat{\Pi}_{n}(t)\delta_{nm},
\end{eqnarray} is satisfied for $n$, $m$ $=1,...,N+1$. The set of projectors in this enlarged
Hilbert space is called the Naimark extension of $\hat{F}_{n}(t)$
\cite{Hel76}. A worked example of this is shown later in this
section.

We now propose that to calculate modal dynamics for POMs, the
preferred observable (determined by the preferred POM) defined by
\erf{POMobservables} becomes a property in ${\cal K}$. The set of
properties which can have definite values is now defined by
\begin{equation}
  {Z}=\{(z_{n}, \hat{\Pi}_n(t))\},
\end{equation} and if $z_n$ is a number this is represented by the operator
  $\hat{Z}=\sum_{n}^{N+1}z_{n}\hat{\Pi}_n(t)$. The value $z_{N+1}$ is arbitrary.
  Note we expect that
  the preferred projective measure $\{\hat\Pi_n(t)\}$ may define more
  then one POM (set of POM elements) as preferred.
The guiding state becomes
\begin{equation} \label{PomState}
  \ket{\Phi(t)}=\ket{\Psi(t)}\otimes\ket{\phi},
\end{equation} where $\ket{\Psi(t)}$ is still the solution to the \sch
equation (\ref{sch}). With this guiding state we can rewrite
\erf{ProbPOM} as
\begin{eqnarray}\label{ProbPOM3}
  P_{n}(t)&=&\bra{\Phi(t)}\hat{\Pi}_{n}(t)\ket{\Phi(t)},
\end{eqnarray} which forces $P_{N+1}(t)\equiv0$ for all time as the projector $\hat{\Pi}_{N+1}(t)$
projects into the null space of $\ket{\Phi(t)}$. The property
states are defined in ${\cal K}$ as
\begin{equation}\label{PomPropertystate}
  \ket{\Phi_{n}(t)}=\hat{\Pi}_{n}(t)\ket{\Phi(t)}/\sqrt{N'},
\end{equation} which now form an orthogonal (distinguishable) set and $N'$ is the new normalization constant. Thus the
standard analysis of modal dynamics now applies (Sec.
\ref{Sec2a}). However, the property state in this extended Hilbert
space is, in general, an entangled state (between the universe and
the auxiliary system). This may raise interpretational
difficulties, but we will leave this question for later work and
just treat the above as a purely mathematical procedure  to deal
with POMs.

To work out the modal dynamics for POMs, as in the projector case,
we need to find the probability current. Using \erf{ProbPOM3} and
\erf{sch}, $J_{nm}(t)$ is
\begin{eqnarray}
J_{nm}(t)&=&2{\rm Im}\{\bra{\Phi(t)}\hat{\Pi}_{n}(t)[\hat{H}_{\rm
uni }(t)\otimes\hat{1}_{\rm aux}-\hat{
R}'(t)]\nl\times\hat{\Pi}_{m}(t)\ket{\Phi(t)}\},
\end{eqnarray} where $\hat{
R}'(t)$ is Hermitian and defines the evolution of the projectors
$\hat\Pi_{n}(t)$, by
\begin{equation}
  d_{t}\hat{\Pi}_{n}=-i[\hat{
R}'(t),\hat{\Pi}_{n}].
\end{equation}

To illustrate the above we consider a simple example, a universe
consisting of only a spin-$1/2$ system \cite{Hel76}, and consider
the observable ${Z}$ defined by the POM elements $
\hat{F}_{n}=\frac{2}{3}\ket{z_{n}}\bra{z_{n}}$, with $n$ being 1,
2, and 3 and $z_{n}=\exp(i2\pi n/3)$.
 The states $\ket{z_{n}}$ are defined by
\beq \ket{z_{n}} = \frac{1}{\sqrt{2}}\Big{(} z_n\ket{1}+
z^{*}_n\ket{2}\Big{)} , \eeq where $\ket{2}$ and $\ket{1}$ are the
eigenstates of $\hat\sigma_{z}$ (the Pauli spin matrix). In the
Bloch sphere these states all lay in the $x-y$ plane with an
angular separation of $2\pi/3$.

Using Naimark's theorem we extend this 2-dimensional Hilbert space
to a 4-dimensional Hilbert space, where it can be shown (using
\erfs{pro1}{pro2}) that the four projectors are \cite{Hel76}
\begin{widetext}\begin{eqnarray}
\hat{\Pi}_{1}&=&\hat{F}_{1}\otimes\ket{\phi}\bra{\phi}-\sqrt{2}/{3}\ket{z_{1}}\bra{z_{2}}
\otimes\ket{\phi}\bra{\phi'}
-\sqrt{2}/{3}\ket{z_{2}}\bra{z_{1}}\otimes\ket{\phi'}\bra{\phi}+1/{3}\ket{z_{2}}\bra{z_{2}}
\otimes\ket{\phi'}\bra{\phi'},\nn\\
\hat{\Pi}_{2}&=&\hat{F}_{2}\otimes\ket{\phi}\bra{\phi}+\sqrt{2}/{3}\ket{z_{2}}\bra{z_{2}}
\otimes\ket{\phi}\bra{\phi'}
+\sqrt{2}/{3}\ket{z_{2}}\bra{z_{2}}\otimes\ket{\phi'}\bra{\phi}+1/{3}\ket{z_{2}}\bra{z_{2}}
\otimes\ket{\phi'}\bra{\phi'},\nn\\
\hat{\Pi}_{3}&=&\hat{F}_{3}\otimes\ket{\phi}\bra{\phi}+\sqrt{2}/{3}\ket{z_{3}}\bra{z_{2}}
\otimes\ket{\phi}\bra{\phi'}
+\sqrt{2}/{3}\ket{z_{2}}\bra{z_{3}}\otimes\ket{\phi'}\bra{\phi}+1/{3}\ket{z_{2}}\bra{z_{2}}
\otimes\ket{\phi'}\bra{\phi'},\nn\\
\hat{\Pi}_{4}&=&\hat{1}\otimes\ket{\phi'}\bra{\phi'}-\ket{z_{2}}\bra{z_{2}}
\otimes\ket{\phi'}\bra{\phi'}.
\end{eqnarray}\end{widetext}
The two states $\ket{\phi}$ and $\ket{\phi'}$ form a set of
orthogonal basis states of the auxiliary Hilbert space and
$\hat{\Pi}_{4}$ is an added projector needed to complete the set
of projectors.

The modal dynamics for these states is formulated as follows. By
\erf{PomState} and \erf{PomPropertystate} the three possible
property states are $\{ \ket{\Phi_{n}(t)}=
\hat{\Pi}_{n}\ket{\phi_1}\ket{\Psi(t)}/\sqrt{N'}$\}, where
$n=1,2,3$. The fourth projector is not included as for all
possible states it projects into the null space of
$\ket{\Phi(t)}$. The observable ${Z}=\{(z_n,\hat{F}_n)\}$ becomes
the property ${Z}=\{(z_n,\hat{\Pi}_n)\}$ (or $\hat{Z}=\sum_n z_n
\hat{\Pi}_{n}$ since for all $n$, $z_n$ is a number), and for this
example the possible values are; $z_{1}=e^{i2\pi/3}$,
$z_{2}=e^{i4\pi/3}$, $z_{3}=1$. The stochastic evolution between
these values is then found from the transition rates $T_{nm}(t)$,
which in turn require a specification of $J_{nm}(t)$. For
simplicity, and to illustrate the significance of enlarging the
Hilbert space, we assume that there is no evolution, that is
$J_{nm}(t)=T_{nm}(t)=0$. Thus once we first assign a property
state (based on the initial probability distribution) it remains
in this property state for all time. Now we assume that the
initial state of the universe is in one of the three
non-orthogonal states, say $\ket{\Psi(0)}=\ket{z_{1}}$. Unlike in
the projector (orthogonal) case we expect there to be some
probability for the universe to be in any of the 3 possible
property states $ \ket{\Phi_{n}(t)}$. This does occur as in
general $\hat{\Pi}_{n}\ket{\phi}\ket{z_{1}}$ is non-zero for all
$n$.

\section{CONTINUOUS MODAL DYNAMICS} \label{Sec3}

In this section we investigate the continuum limit of modal
dynamics. This has been previously done by Sudbery and Vink in
Refs. \cite{Sud87} and \cite{Vin93} respectively, where it was
shown that Bohmian mechanics can be obtained by choosing the
appropriate preferred projective measure (property states). In
this section we
 briefly outline their work, then extend it to show that the Brown
and Hiley generalization of Bohmian mechanics to include the
 momentum representations \cite{BroHil01} is not the
continuum limit of Bell's modal dynamics. We  also present an
alternative (we believe easier) method for finding the continuous
trajectories, that works when the modal dynamics has a continuum
limit.

\subsection{Bohmian Mechanics}

In Bohmian mechanics \cite{Boh} the preferred projective measure
is the one associated with the position basis. That is, the
property is the position of the system $\{X_{j}\}$ (vector
notation) and the wavefunction
$\Psi(\{x_{j}\},t)=\langle{\{x_{j}\}}\ket{\Psi(t)}$ is then
interpreted as a field which guides the position in a
non-classical way. In Bohm's original papers he showed that this
non-classical behavior could be represent by an extra potential in
the Hamiltonian-Jacobi equation, the quantum potential which
depends on $\Psi(\{x_{j}\},t)$.

In this paper we do not introduce the quantum potential, but
instead describe Bohm trajectories with reference to a continuous
probability current $J_{k}(\{x_{j}\},t)$. We also consider
$\ket{\Psi(t)}$ to belong to a tensor product ${\cal
H}_{x}\otimes{\cal H}'$, where ${\cal H}_{x}$ is the Hilbert space
containing the position eigenstates $\ket{\{x_j\}}$. Then the
wavefunction becomes a vector given by,
\begin{equation} \label{condState}
\ket{\tilde{\psi}(\{x_{j}\},t)}=\langle{\{x_{j}\}}\ket{\Psi(t)}\in{\cal
H}'.
\end{equation} This allows us to take into account a Hilbert
space for the universe which is larger than that for the position
of the system. Here we see that $
\ket{\{x_{j}\}}\ket{\tilde{\psi}(\{x_{j}\},t)}$ is the continuous
equivalent of our unnormalized property state, and the properties
are the position operators $\{\hat{X}_{k}\}$. With this property
state we can define a probability density as
\begin{equation}\label{probDen}
 P(\{x_{j}\},t)=\langle{\tilde{\psi}(\{x_{j}\},t)}\ket{\tilde{\psi}(\{x_{j}\},t)}
\end{equation}
which obeys the continuity equation
\begin{eqnarray}\label{continuity}
\partial_{t}
P(\{x_{j}\},t)&=&-\sum_{k}\partial_{x_{k}}J_{k}(\{x_{j}\},t),
\end{eqnarray} where $J_{k}(\{x_{j}\},t)$ depends on the form of
$H_{\rm uni}(t)$. As in the modal case there is not a unique
solution to $J_{k}(\{x_{j}\},t)$ as for example in three
dimensions we can add any vector field $\nabla\times{\bf
A}(\{x_{j}\},t)$ to ${\bf J}(\{x_{j}\},t)$ as
$\nabla\cdot\nabla\times{\bf A}(\{x_{j}\},t)\equiv0$.

We define a velocity field $v_{k}(\{x_{j}\},t)$ implicitly by
\begin{equation} \label{Jinvel}
  J_{k}(\{x_{j}\},t)=v_{k}(\{x_{j}\},t) P(\{x_{j}\},t).
\end{equation}
Bohmian trajectories for $\{x_{j}\}(t)$ are then defined by
\begin{equation}\label{diff}
  d_{t}x_{k}(t)=v_{k}(\{x_{j}\},t)|_{x_{k}=x_{k}(t)}.
\end{equation} Probability enters only through the initial conditions,
$\{x_{k}(0)\}$. This is only one of the possible continuous
trajectories which satisfy \erf{probDen}. Other possibilities
include stochastic approaches; see Ref. \cite{Vin93} and
references within.
It should be noted that so far nothing has specified that
$\{x_{j}\}$ must be positions, and in fact Brown and Hiley
\cite{BroHil01} develop a formalism
 where $\{x_{j}\}$ can be either position or momentum.
 As examples, they consider a simple universe (a single 1-D particle) and
derive $d_{t}x(t)$ and $d_{t}p(t)$ for a linear, quadratic and
cubic potential.

\subsubsection{Bohmian mechanics as the continuous limit of discrete modal dynamics}

To demonstrate that the modal dynamics does give Bohmian mechanics
as its continuum limit, first consider the Hamiltonian
\begin{eqnarray} \label{generalHam}
\hat{H}_{\rm
uni}(t)&=&\hat{A}(t)+\sum_{k}\hat{B}_{k}(t)\hat{Y}_{k}+\sum_{k}\hat{Y}_{k}\hat{B}_{k}\dg(t) \nl
+\sum_{k}\hat{C}_{k}(t)\hat{Y}^{2}_{k}+\sum_{k}\hat{Y}^{2}_{k}\hat{C}_{k}\dg(t),
\end{eqnarray} where $\hat{A}(t)$, $\hat{B}_{k}(t)$ and $\hat{C}_{k}(t)$ are arbitrary functions of
the operators $\{\hat{X}_{k}\}$ and the rest of the universe, and
$\{\hat{Y}_k\}$ are the conjugate operators to $\{\hat{X}_{k}\}$.
That is, $[\hat Y_j,\hat X_k]=-i\delta_{jk}$.

To calculate $d_t{x}_{k}(t)$ in \erf{diff} we need to calculate
the velocity field $v_k(\{x_{j}\},t)$, which in turn requires
calculation of $J_{k}(\{x_{j}\},t)$. Taking the derivative of
\erf{probDen} and using the \sch equation (\erf{sch}) we can write
\begin{widetext}\begin{eqnarray}
  d_{t}
P(\{x_{j}\},t)&=&2{\rm
Im}[\bra{{\Psi}(t)}\{x_{j}\}\rangle\bra{\{x_{j}\}}\hat{H}_{\rm
uni}(t)\ket{{\Psi}(t)}] \nn \\
&=&-2\sum_{k}{\rm Re}\Big{[} \bra{\tilde{\psi}(\{x_{j}\},t)}\{
2{\rm Re}[\hat{B}_{k}(\{x_{j}\},t)]\partial_{x_{k}} +
\partial_{x_{k}}{\rm Re}[\hat{B}_{k}(\{x_{j}\},t)]\}
 \ket{\tilde{\psi}(\{x_{j}\},t)}\Big{]} \nl
-2\sum_{k}{\rm Im}\Big{[} \bra{\tilde{\psi}(\{x_{j}\},t)}\{ 2{\rm
Re}[\hat{C}_k(\{x_{j}\},t)]\partial_{x_{k}}^{2} +2
\partial_{x_{k}}\hat{C}^{\dagger}_k(\{x_{j}\},t)\partial_{x_{k}}+
\partial_{x_{k}}^{2}\hat{C}^{\dagger}_k(\{x_{j}\},t)
  \}
 \ket{\tilde{\psi}(\{x_{j}\},t)}\Big{]},\nl
\end{eqnarray} which can be shown to be
\begin{equation}
  d_{t}
P(\{x_{j}\},t)=-\sum_{k}\partial_{x_k}\Big{\{} {\rm
Re}\Big{[}\bra{\tilde{\psi}(\{x_{j}\},t)}\{2{\rm Re}[
\hat{B}_{k}(\{x_{j}\},t)]-4i{\rm Re}
[\hat{C}_k(\{x_j\},t)]\partial_{x_{k}}-2i\partial_{x_{k}}[\hat{C}\dg_k(\{x_j\},t)]\}
\ket{\tilde{\psi}(\{x_{j}\},t)}\Big{]}\Big{\}}.
\end{equation}
Comparing this with \erf{continuity} gives
\begin{eqnarray} \label{generalJ}
J_{k}(\{x_{j}\},t)&=&{\rm
Re}\Big{[}\bra{\tilde{\psi}(\{x_{j}\},t)}\{2{\rm Re}[
\hat{B}_{k}(\{x_{j}\},t)]-4i{\rm Re}
[\hat{C}_k(\{x_j\},t)]\partial_{x_{k}}-2i\partial_{x_{k}}[\hat{C}\dg_k(\{x_j\},t)]\}
\ket{\tilde{\psi}(\{x_{j}\},t)}\Big{]}.
\end{eqnarray}\end{widetext}

For simplicity we consider the case when
$\hat{A}(t)=\hat{V}(\hat{X})$, $\hat{B}_{k}(t)=0$, and
$\hat{C}(t)=\hat{C}^{\dagger}(t)=\hat{1}/4M$, describing for
example
 an electron in a 1-D potential, with
the operator nature of $\hat{1}$ and $\hat{V}$ signifying
operation on the Hilbert space for the internal structure
of the electron. } For this example $J(x,t)$ becomes
\begin{equation}
J(x,t)=\frac{1}{M} {\rm
Im}[\bra{\tilde{\psi}(x,t)}\partial_{x}\ket{\tilde{\psi}(x,t)}],
\end{equation} and thus
\begin{equation}
  d_{t}x(t)=\frac{1}{M}{\rm Im}\Big{[} \frac{\langle{\tilde{\psi}(x,t)}|
  \partial_{x_{k}}
  \ket{\tilde{\psi}(x,t)}}{\langle{\tilde{\psi}(x,t)}\ket{\tilde{\psi}(x,t)}}\Big{]}\Big{|}_{x=x(t)}.
\end{equation}
To simplify this we can rewrite $\ket{\tilde{\psi}(x,t)}$ as
\begin{equation}
  \ket{\tilde{\psi}(x,t)}=\sum_{j}R_{j}(x,t)\exp[iS_{j}(x,t)]\ket{j}
\end{equation} where $R_{j}(x,t)$ and $S_{j}(x,t)$ are real
functions and $\{\ket{j}\}$ is an orthonormal basis set,
which for example spans the Hilbert space of the internal structure of the electron. Then it
simplifies to
\begin{equation} \label{conlimit1}
  d_{t}x(t)=
  \frac{\sum_{j}R^{2}_{j}(x,t)\partial_{x}[S_{j}(x,t)]}{M\sum_{j}R_{j}^{2}(x,t)}\Big{|}_{x=x(t)}.
\end{equation}

To compare this to the modal dynamics defined in Sec. \ref{Sec2}
 we have to discretize $X$. In
\cite{Vin93} and \cite{Sud87} this is done by defining a lattice
of size $N$ and lattice separation $\epsilon$. Thus the values of
the property $X$ denoted $x$ become $x_{n}=\epsilon n$, and the
preferred projective measure becomes
$\{\hat{\pi}_{n}=\ket{\epsilon n}\bra{\epsilon n}\otimes
\hat{1}\}$. With this preferred projective measure the property
state \erf{propertyState} becomes $\ket{\tilde{\Psi}_{n}(t)}=\ket{
\epsilon n}\bra{\epsilon n}{\Psi(t)}\rangle=\ket{\epsilon
n}|{\tilde{\psi}_{n}(t)}\rangle$ where
$|{\tilde{\psi}_{n}(t)}\rangle$ is an unnormalized state existing
in ${\cal H}'$. Using the results of Sec. \ref{Sec2a} the
probability current is
\begin{equation} \label{Jnmx}
  J_{nm}(t)=2{\rm Im}[\bra{\tilde{\psi}_{n}(t)}\bra{\epsilon n}\hat{H}_{\rm uni
}\ket{\epsilon m}\ket{\tilde{\psi}_{m}(t)}],
\end{equation}
and the discretized version of the Hamiltonian is
\begin{eqnarray}
\bra{\epsilon n}\hat{H}_{\rm uni }\ket{\epsilon m}&=&
-(\delta_{n,m+1}+\delta_{n,m-1}-2\delta_{n,m})/2M\epsilon^2\nl+\hat{V}(\epsilon
n)\delta_{n,m}.
\end{eqnarray} This gives
\begin{eqnarray}
  J_{nm}(t)&=&\frac{-1}{M\epsilon^{2}}{\rm Im}[\bra{\tilde{\psi}_{n}(t)}{\tilde{\psi}_{n-1}(t)}
  \rangle\delta_{n,m+1}\nl+
  \bra{\tilde{\psi}_{n}(t)}{\tilde{\psi}_{n+1}(t)}\rangle\delta_{n,m-1}].
\end{eqnarray}
Taylor expanding $|{\tilde{\psi}_{n+1}(t)}\rangle$ and
$|{\tilde{\psi}_{n-1}(t)}\rangle$ gives
\begin{eqnarray}
  J_{nm}(t)&=&\frac{1}{M\epsilon}{\rm
  Im}\{\bra{\tilde{\psi}_{n}(t)}\Delta_\epsilon[|{\tilde{\psi}_{n}(t)}\rangle]\delta_{n,m+1}\nl-
  \bra{\psi_{n}(t)}\Delta_\epsilon[|{\tilde{\psi}_{n}(t)}\rangle]\delta_{n,m-1}+O(\epsilon)\},\hspace{1cm}
\end{eqnarray} where $\Delta_\epsilon$ is the discretized version of a
derivative. As in the continuous case we write
$|{\tilde{\psi}_{n}(t)}\rangle$ in terms of the real functions
$S_{j}(\epsilon n,t)$ and $R_{j}(\epsilon n,t)$ which results in
$J_{nm}(t)$ becoming
\begin{eqnarray}
  J_{nm}(t)&=&\frac{1}{M\epsilon }\{\sum_{j}R_{j}^{2}(\epsilon n)\Delta_\epsilon[S_{j}(\epsilon n)]\delta_{n,m+1}\nl-
  \sum_{j}R_{j}^{2}(\epsilon n)\Delta_\epsilon[S_{j}(\epsilon n)]\delta_{n,m-1}+O(\epsilon)\}.
  \nl
\end{eqnarray}
Since $1/\epsilon \gg 1$, in the $\epsilon\rightarrow 0$ limit
(continuum limit) we can neglect the higher order terms in the
above expression for $J_{nm}(t)$. That is the only terms which
will contribute are the transitions from $m$ to $m-1$ or $m$ to
$m+1$. Because of this we can write
\begin{eqnarray}
  J_{n(n+1)}(t)&=-&\frac{1}{M\epsilon}\sum_{j}R_{j}^{2}(\epsilon n)\Delta_\epsilon [S_{j}(\epsilon n)]\\
    J_{n(n-1)}(t)&=&\frac{1}{M\epsilon}\sum_{j}R_{j}^{2}(\epsilon n)\Delta_\epsilon[S_{j}(\epsilon n)].
\end{eqnarray}
If $\sum_{j}R_{j}^{2}\Delta_\epsilon [S_{j}(\epsilon n)]>0$
(implies $J_{n(n+1)}(t)<0$ and $J_{n(n-1)}(t)>0$) then by
\erft{t1}{t4},
\begin{eqnarray}
  T_{(n+1)n}(t)&=&\frac{\sum_{j}R_{j}^{2}(\epsilon n)\Delta_\epsilon[S_{j}
  (\epsilon n)]}{M\epsilon \sum_{j}R_{j}^{2}(\epsilon n)}\\
  T_{(n-1)n}(t)&=&0.
  \end{eqnarray}
If $\sum_{j}R_{j}^{2}\Delta_\epsilon[S_{j}(\epsilon n)]<0$
(implies $J_{n(n+1)}(t)>0$ and $J_{n(n-1)}(t)<0$) then by
\erft{t1}{t4}
\begin{eqnarray}
  T_{(n+1)n}(t)&=&0\\
  T_{(n-1)n}(t)&=&-\frac{\sum_{j}R_{j}^{2}(\epsilon n)\Delta_\epsilon [S_{j}
  (\epsilon n)]}{M\epsilon\sum_{j}R_{j}^{2}(\epsilon n)}.
  \end{eqnarray}
These transition rates imply that in an interval $dt$ the average
displacement $dx$ will be
\begin{eqnarray}
  E[dx]&=&\epsilon T_{(n+1)n}dt-\epsilon T_{(n-1)n}dt\nn\\&=&
  \frac{\sum_{j}R_{j}^{2}(\epsilon n)\Delta_\epsilon[S_{j}(\epsilon n)]}{M\sum_{j}R_{j}^{2}(\epsilon
  n)}dt+O(\epsilon)dt.\hspace{1cm}
\end{eqnarray}
Provided $S_{j}(\epsilon n)$ and $R_{j}(\epsilon n)$ are
continuous, the average $E[dx]$ reduces to \erf{conlimit1} as
$\epsilon\rightarrow 0$. However, to show that the trajectories
are smooth and deterministic from the initial $\{x_{k}(0)\}$ we
also require that the dispersion $E[dx^{2}]$ goes to zero in the
continuum limit. This is the case since
\begin{equation}
  E[dx^{2}]=\epsilon^{2} T_{(n+1)n}+\epsilon^{2}dt
  T_{(n-1)n}dt=O(\epsilon)dt,
\end{equation} which  goes to zero as $\epsilon\rightarrow0$.

\subsubsection{Hamiltonians for which Bohmian mechanics is not the continuous limit of discrete modal dynamics}

The above demonstrates that in the continuum limit modal dynamics
becomes Bohmian Mechanics. However, if we consider a Hamiltonian
of the form
\begin{equation}
  H_{\rm uni}=\kappa\hat{Y}^{3}+V(\hat{X}),
\end{equation} we find that the continuous limit of discrete (Bell-type) modal
dynamics is not Bohmian mechanics.
 Note this Hamiltonian is unreasonable if $\hat{X}$ is position as
this says we have a cubic dependence on momentum, which is not
present in any natural Hamiltonians. However, if $\hat{X}$
corresponds to momentum and $-\hat{Y}$ to position, (which occurs
in in Brown and Hiley's \cite{BroHil01} extension of Bohmian
mechanics to include the momentum representation), then this
Hamiltonian is valid.

As before if we discretize the value of $X$, $x\rightarrow
x_{n}=\epsilon n$, the probability current again will be given by
\erf{Jnmx}. However, the discretized version of the Hamiltonian in
this case is
\begin{eqnarray}
\bra{\epsilon n}\hat{H}_{\rm uni }\ket{\epsilon
m}&=&\frac{\kappa}{\epsilon^{3}}(i\delta_{n,m+3}-i\delta_{n,m-3}-3i\delta_{n,m+1}\nl+3i\delta_{n,m-1})
+V(\epsilon
n)\delta_{n,m},
\end{eqnarray} which results in
\begin{widetext}\begin{eqnarray}
  J_{nm}(t)&=&\frac{\kappa}{\epsilon^{3}}{\rm Im}\Big{[}i\delta_{n,m+3}\bra{\tilde{\psi}_{n}(t)}
  {\tilde{\psi}_{n-3}(t)}\rangle
  -i\delta_{n,m-3}\bra{\tilde{\psi}_{n}(t)}{\tilde{\psi}_{n+3}(t)}\rangle
  -3i\delta_{n,m+1}\bra{\tilde{\psi}_{n}(t)}{\tilde{\psi}_{n-1}(t)}\rangle
   \nl+3i\delta_{n,m-1}\bra{\tilde{\psi}_{n}(t)}{\tilde{\psi}_{n+1}(t)}\rangle\Big{
   ]}.
\end{eqnarray}\end{widetext} Taylor expanding this gives a rather large expression, but since
 $\bra{\tilde{\psi}_{n}(t)}{\tilde{\psi}_{n}(t)}\rangle$
is a real number and $1/\epsilon^{3}\gg1/\epsilon^{2}$ in the
$\epsilon\rightarrow0$ limit we can ignore all orders of the
Taylor expansion. This allows us to write
\begin{eqnarray}
  J_{nm}(t)&=&\frac{\kappa\bra{\tilde{\psi}_{n}(t)}{\tilde{\psi}_{n}(t)}\rangle}{\epsilon^{3}}[\delta_{n,m+3}-\delta_{n,m-3}-3
  \delta_{n,m+1}\nl+3\delta_{n,m-1}].
\end{eqnarray}
Since $\bra{\tilde{\psi}_{n}(t)}{\tilde{\psi}_{n}(t)}\rangle$ is
always positive, for $\kappa>0$ the transitions rates defined in
\erft{t1}{t4} become
\begin{eqnarray}
T_{(n-3)n}(t)&=&0\\
T_{(n+3)n}(t)&=&\frac{\kappa}{\epsilon^{3}}\\
T_{(n-1)n}(t)&=&\frac{3\kappa}{\epsilon^{3}}\\
T_{(n+1)n}(t)&=&0.
\end{eqnarray}

These transition rates imply that in an interval $dt$ the average
displacement $dx$ will be
\begin{eqnarray}
  M[dx]&=&3\epsilon T_{(n+3)n}dt-\epsilon T_{(n-1)n}dt=0
\end{eqnarray}
and the dispersion will be
\begin{equation}
  M[dx^{2}]=9\epsilon^{2} T_{(n+1)n}+\epsilon^{2}dt T_{(n-1)n}dt=12 \kappa dt/\epsilon
\end{equation} which diverges as $\epsilon\rightarrow0$. Thus the
continuum limit does not exists.  This implies that Brown and
Hiley's \cite{BroHil01} extension of Bohmian mechanics to include
the momentum representation (the momentum projector is the
preferred projective measure) is not the continuum limit of Bell's
modal dynamics. It is possible that a different choice for
$J_{nm}(t)$ (and $T_{nm}(t)$) would allow their equations to be
derived, but that is beyond the scope of this paper.

\subsection{The Velocity Operator Technique} \label{Sec:vel}

In the above section we have demonstrated that when using the Bell
solution for the transition rates, modal dynamics for some
continuous properties only reduces to a deterministic theory
(apart from a random initial conditions) if the Hamiltonian is at
most quadratic in the conjugate variable to the property. If the
property is position then this deterministic limit is Bohmian
mechanics and the trajectories are then found using \erf{diff},
which requires calculation of the probability current density,
$J_k(\{x_{j}\},t)$. Here we present an alternative to this, a
method to calculate $v_k(\{x_{j}\},t)$ directly. We assume that
\begin{equation} \label{velocityField}
  v_{k}(\{x_{j}\},t)=\frac{{\rm Re}[\bra{\Psi(t)}{\{x_{j}\}}\rangle\bra{\{x_{j}\}}\hat{v}_{k}(t)\ket{\Psi(t)}]}
  {\bra{\Psi(t)}{\{x_{j}\}}\rangle\bra{\{x_{j}\}}\Psi(t)\rangle},
\end{equation}
where $\hat{v}_{k}(t)$ is the $k^{\rm th}$ component of the
velocity operator. This operator is defined as
\begin{equation} \label{velocityO}
  \hat{v}_{k}(t)=-i[\hat{X}_k,\hat{H}_{\rm uni}(t)].
\end{equation}

To show that this does give the same trajectories as Bohmian
mechanics, we note that the numerator of \erf{velocityField}
should be $J_{k}(\{x_{j}\},t)$ by definition. Now using the
Hamiltonian defined in \erf{generalHam}, the velocity operator is
\begin{equation}
 \hat{v}_{k}(t)=\hat{B}_{k}(t)+\hat{B}_{k}\dg(t)+2
\hat{C}_k(t)\hat{Y}_{k}+2\hat{Y}_{k}\hat{C}\dg_k(t).
\end{equation} This results in the following velocity field
\begin{widetext}\begin{eqnarray}
v_{k}(\{x_{j}\},t)&=&
\frac{1}{\bra{\tilde{\psi}(\{x_{j}\},t)}\tilde{\psi}(\{x_{j}\},t)\rangle}{\rm
Re}\Big{[}\bra{\tilde{\psi}(\{x_{j}\},t)}\{2{\rm Re}[
\hat{B}_{k}(\{x_{j}\},t)]-4i{\rm Re}
[\hat{C}_k(\{x_j\},t)]\partial_{x_{k}}\nl-2i\partial_{x_{k}}[\hat{C}\dg_k(\{x_j\},t)]\}
\ket{\tilde{\psi}(\{x_{j}\},t)}\Big{]}.
\end{eqnarray}\end{widetext} Comparing this with \erf{generalJ} we see that the
numerator is indeed $J_{k}(\{x_{j}\},t)$. This completes our proof
that our velocity method does generate the same trajectories as
Bohmian mechanics for Hamiltonians of the form displayed in
\erf{generalHam}. However, by extending this argument to higher
orders it can shown that our velocity method does not agree with
Bohmian mechanics. That is, this is another example of a method
that only works for Hamiltonians that do not contain terms of
order $\hat Y_{k}^3$ and higher.

\section{SIMPLE EXAMPLE - HARMONIC OSCILLATOR} \label{Sec4}

\subsection{Husimi POM} \label{Husimi}

To illustrate modal dynamics for POMs, we investigate a simple
model; a universe consisting of a one dimensional harmonic
oscillator of frequency $\omega$. That is,
\begin{equation}\label{HuniExample}
  \hat{H}_{\rm uni}(t)=\omega \hat{a}\dg\hat{a},
\end{equation} where $\hat{a}\dg$ and $\hat{a}$ are the
creation and annihilation operators of the harmonic oscillator
respectively.

The preferred POM we consider is that of Ref. \cite{Husimi},
which has POM elements (or effects) given by
\begin{equation}\label{POMexample}
  \hat{F}_{\alpha}=\frac{1}{\pi}\ket{\alpha}\bra{\alpha}d^2\alpha,
\end{equation}
where $\hat{a}\ket{\alpha}=\alpha\ket{\alpha}$.
 The preferred observable (one
of the many) we associate with this POM is
\begin{equation}\label{ObservableExample}
  {A}=\{(\alpha,\frac{1}{\pi}\ket{\alpha}\bra{\alpha}d^2\alpha)\}
\end{equation}
 The continuous value $\alpha=x^{+}+iy^{-}$ is a complex number
 representing a point is phase space ($x^{+}$,$y^{-}$). In the orthodox theory this POM corresponds to a measure
of both position and momentum with minimal additional uncertainty
\cite{Hel76}.

Before analyzing the modal dynamics that corresponds to this POM,
we need to define a few operators that act in the enlarged Hilbert
space ${\cal K}$. (The auxiliary Hilbert space ${\cal H}_{\rm
aux}$ is assumed to be a single harmonic oscillator). We define
\begin{eqnarray}\label{x+y-etc}
\hat{x}^{+}&=&[\hat{a}+\hat{a}\dg+\hat{b}+\hat{b}\dg]/2, \\
\hat{x}^{-}&=&[\hat{a}+\hat{a}\dg-\hat{b}-\hat{b}\dg]/2, \\
\hat{y}^{+}&=&[-i\hat{a}+i\hat{a}\dg-i\hat{b}+i\hat{b}\dg]/2, \\
\hat{y}^{-}&=&[-i\hat{a}+i\hat{a}\dg+i\hat{b}-i\hat{b}\dg]/2,
\end{eqnarray} where
$\hat{b}$ and $\hat{b}\dg$ are annihilation and creation operators
which act in ${\cal H}_{\rm aux}$. These four operators obey the
commutator relations
\begin{eqnarray}\label{commutators}
&&[\hat{x}^{+},\hat{y}^{+}]=[\hat{x}^{-},\hat{y}^{-}]=i, \\
&&[\hat{x}^{+},\hat{y}^{-}]=[\hat{x}^{-},\hat{y}^{+}]=0,
\end{eqnarray} thus $\hat{x}^{+}$ and $\hat{y}^-$  have the same eigenstates,
which we denote  as $\ket{x^{+},y^{-}}$. They are given by
 \begin{equation}\label{x+y-state}
\ket{x^{+},y^{-}}=\int
\frac{dx'}{\sqrt{2\pi}}\Ket{\frac{x^{+}-x'}{\sqrt{2}}}_{\rm
aux}\Ket{\frac{x^{+}+x'}{\sqrt{2}}}_{\rm uni}e^{iy^{-}x'},
\end{equation} where $\ket{{(x^{+}+x')}/{\sqrt{2}}}_{\rm uni}$ is
an $x$-state (an eigenstate of
$\hat{X}=(\hat{a}+\hat{a}\dg)/\sqrt{2}$).

Because  $\hat{x}^{+}$ and $\hat{y}^{-}$ can be well-defined
simultaneously, we interpreted them as being suitable modal properties to represent
 simultaneously the
position and momentum of the harmonic oscillator.  This can be justified  on
the grounds that for $\ket{\phi}=\ket{0}$ (a vacuum state)
\begin{eqnarray}
  \bra{\Phi(t)}\hat x^{+}\ket{\Phi(t)}&=&\bra{\Psi(t)}\hat{X}
  \ket{\Psi(t)}/\sqrt{2},\\
   \bra{\Phi(t)}\hat y^{-}\ket{\Phi(t)}&=&\bra{\Psi(t)}\hat{Y}
  \ket{\Psi(t)}/\sqrt{2},\\
   \bra{\Phi(t)}{\hat x^{+}}{\hat x^{+}}\ket{\Phi(t)}&=&\bra{\Psi(t)}\hat{X}^{2}
  \ket{\Psi(t)}/2+1/4,\\
   \bra{\Phi(t)}{\hat y^{-}}{\hat y^{-}}\ket{\Phi(t)}&=&\bra{\Psi(t)}\hat{Y}^{2}
  \ket{\Psi(t)}/2+1/4, \hspace{1cm}
\end{eqnarray} where $\ket{\Phi(t)}=\ket{\Psi(t)}\ket{0}$ is the
guiding wave for ${\cal K}$ and $\ket{\Psi(t)}$ is the solution of
the \sch equation. Thus  these operators have essentially the same
statistics as the position ($\hat{X}$) and momentum
($\hat{Y}=(-i\hat{a}+i\hat{a}\dg)/\sqrt{2}$) operators in the
classical limit (as the 1/4 term becomes negligible).

In ${\cal K}$ we can  rewrite $\hat{H}_{\rm uni}$ as
\begin{eqnarray}\label{Huniext}
  \hat{H}_{\rm uni}\otimes\hat{1}_{\rm aux}&=&\omega[{\hat{x}^{+}}\hat{x}^{+}+{\hat{x}^{-}}\hat{x}^{-}
  +{\hat{y}^{+}}\hat{y}^{+}
  +{\hat{y}^{-}}\hat{y}^{-}\nl
  +2\hat x^{+}\hat{x}^{-}+2\hat{y}^{+}\hat{y}^{-}-2]/4.
\end{eqnarray}
To define the modal dynamics in ${\cal K}$ we use Naimark theorem
and a Naimark projector $\ket{\phi}=\ket{0}$ to extend the POM
elements defined by \erf{POMexample} to the projector
$\ket{x^{+},y^{-}}\bra{x^{+},y^{-}}dx^{+}dy^{-}$. That is
\begin{equation} \label{NiaExample1}
  \frac{1}{\pi}\ket{\alpha}\bra{\alpha}d^2\alpha=\langle{0}\ket{x^{+},y^{-}}\bra{x^{+},y^{-}}0\rangle dx^{+}dy^{-}
\end{equation} (see appendix \ref{A}).
 The observable
$A$ becomes a property, which in matrix notation is given by
\begin{equation}\label{propertyExample}
  \hat{A}=\int\int (x^{+}+iy^{-}) \ket{x^{+},y^{-}}\bra{x^{+},y^{-}}dx^{+}
  dy^{-}.
\end{equation} Since the preferred projective measure in ${\cal K}$,
$\{\hat\Pi(x^{+},y^{-})dx^{+}dy^{-}=\ket{x^{+},y^{-}}\bra{x^{+},y^{-}}dx^{+}dy^{-}\}$
forms a complete orthogonal set and \erf{Huniext} contains no
cubic or higher order terms involving $\hat x^{-}$ or
$\hat{y}^{+}$, the results of Sec. \ref{Sec:vel} are applicable to
this paper. That is a deterministic differential equation for the
values $x^{+}(t)$ and ${y}^{-}(t)$ can be determined.

Using \erf{velocityO} with the Hamiltonian \erf{Huniext} gives the
following two velocity operators
\begin{eqnarray}\label{VelocityOexa}
  \hat{v}_{+}(t)&=&\frac{\omega}{2}\hat{y}^{+}+\frac{\omega}{2}\hat{y}^{-},\\
    \hat{v}_{-}(t)&=&-\frac{\omega}{2}\hat{x}^{+}-\frac{\omega}{2}\hat{x}^{-}.
\end{eqnarray} Substituting these into \erf{velocityField} (with $\ket{\Psi(t)}\rightarrow\ket{\Phi(t)}$)
gives
\begin{widetext}
\begin{eqnarray}\label{Velocityfieldexa}
{v}_{+}(x^{+},y^{-},t)&=&\frac{\omega{\rm Re}[-i
\langle\Phi(t)\ket{x^{+},y^{-}}\partial_{x^{+}}
\bra{x^{+},y^{-}}\Phi(t)\rangle]}{2
\langle\Phi(t)\ket{x^{+},y^{-}}\bra{x^{+},y^{-}}\Phi(t)\rangle}+\frac{\omega}{2}{y^{-}},\nl\\
{v}_{-}(x^{+},y^{-},t)&=&-\frac{\omega{\rm Re}[i
\langle\Phi(t)\ket{x^{+},y^{-}}\partial_{y^{-}}
\bra{x^{+},y^{-}}\Phi(t)\rangle]}{2
\langle\Phi(t)\ket{x^{+},y^{-}}\bra{x^{+},y^{-}}\Phi(t)\rangle}-\frac{\omega}{2}{x^{+}}.\nl
\end{eqnarray}
Thus the differential equations are
\begin{eqnarray}\label{diffgenexa}
 d_t{x}^{+}(t)&=&\frac{\omega{\rm Re}[-i
\langle\Phi(t)\ket{x^{+},y^{-}}\partial_{x^{+}}
\bra{x^{+},y^{-}}\Phi(t)\rangle]}{2
\langle\Phi(t)\ket{x^{+},y^{-}}\bra{x^{+},y^{-}}\Phi(t)\rangle}\Big{|}_{x^{+}=x^{+}(t),y^{-}=y^{-}(t)}
+\frac{\omega}{2}{y^{-}}(t),\\
 d_t{y}^{-}(t)&=&-\frac{\omega{\rm Re}[i
\langle\Phi(t)\ket{x^{+},y^{-}}\partial_{y^{-}}
\bra{x^{+},y^{-}}\Phi(t)\rangle]}{2
\langle\Phi(t)\ket{x^{+},y^{-}}\bra{x^{+},y^{-}}\Phi(t)\rangle}\Big{|}_{x^{+}=x^{+}(t),y^{-}=y^{-}(t)}
-\frac{\omega}{2}{x^{+}}(t).
\end{eqnarray}
Using the fact that (see appendix \ref{A})
\begin{equation}
\bra{x^{+},y^{-}}\Phi(t)\rangle
=\frac{\exp[-({x^{+}}^{2}+{y^{-}}^{2})/2]}{\sqrt{\pi}}\sum_{m}
\frac{(x^{+}-iy^{-})^{m}}{\sqrt{m!}}\bra{m}{\Psi(t)}\rangle
\end{equation} the partial derivatives can be written as
\begin{eqnarray}
\partial_{x^{+}} \bra{x^{+},y^{-}}\Phi(t)\rangle &=& \frac{\exp
[-({x^{+}}^{2}+{y^{-}}^{2})/2]}{\sqrt{\pi}}\sum_{m}m
\frac{(x^{+}-iy^{-})^{m-1}}{\sqrt{m!}}\bra{m}{\Psi(t)}\rangle
-x^{+}\bra{x^{+},y^{-}}\Phi(t)\rangle
,\\
\partial_{y^{-}}\bra{x^{+},y^{-}}\Phi(t)\rangle &=&-i\frac{\exp
[-({x^{+}}^{2}+{y^{-}}^{2})/2]}{\sqrt{\pi}}\sum_{m}m
\frac{(x^{+}-iy^{-})^{m-1}}{\sqrt{m!}}\bra{m}{\Psi(t)}\rangle
-y^{-}\bra{x^{+},y^{-}}\Phi(t)\rangle.
\end{eqnarray}\end{widetext}
which allows us to write
\begin{eqnarray}\label{diffgenexa2}
 d_t{x}^{+}(t)&=&+\frac{\omega}{2}{y^{-}}(t)+\frac{\omega}{2}{\rm Im}[\chi_{\psi}(x^{+}(t),y^{-}(t))],\hspace{1cm}\\
 d_t{y}^{-}(t)&=&-\frac{\omega}{2}{x^{+}}(t)-\frac{\omega}{2}{\rm
Re}[\chi_{\psi}(x^{+}(t),y^{-}(t))],
\end{eqnarray}
where
\begin{equation}\label{chi}
  \chi_{\psi}(x^{+}(t),y^{-}(t))=\frac{\sum_{m}m{(x^{+}-iy^{-})^{m-1}}\bra{m}{\Psi(t)}\rangle/{\sqrt{m!}}}
{\sum_{m}
{(x^{+}-iy^{-})^{m}\bra{m}{\Psi(t)}\rangle}/{\sqrt{m!}}}.
\end{equation}
Thus the differential equation for $\alpha(t)$ is
\begin{equation}\label{alphat}
   d_t{\alpha}(t)=-\frac{i\omega}{2}\alpha(t)-\frac{i\omega}{2}\chi_{\psi}(x^{+}(t),y^{-}(t))
\end{equation}

\subsubsection{When $\ket{\Psi(t)}$ is a number state}

Lets first of all consider a number state $\ket{n}$, as the
initial condition for $\ket{\Psi(0)}$. Then by the \sch equation
\begin{equation}\label{schExample}
  d_{t}\ket{\Psi(t)}=-i\omega \hat{a}\dg\hat{a}\ket{\Psi(t)},
\end{equation} $\ket{\Psi(t)}=e^{-i\omega n t}\ket{n}$. Substituting this
into \erf{chi}, gives
\begin{eqnarray}\label{chinumber}
  \chi_{\psi}(x^{+}(t),y^{-}(t))&=&\frac{n{[x^{+}(t)-iy^{-}(t)]^{n-1}}}
{
[x^{+}(t)-iy^{-}(t)]^{n}}=\frac{n}{\alpha^{*}(t)}\nl=\frac{n\alpha(t)}{|\alpha(t)|^{2}}.
\end{eqnarray}
Thus
\begin{equation}
   d_t{\alpha}(t)=-\frac{i\omega}{2}\Big{(}1+\frac{n}{|\alpha(t)|^{2}}\Big{)}\alpha(t).
\end{equation}
This has the solution
\begin{equation}\label{alphanum}
  \alpha(t)=\alpha(0)e^{-i \omega' t},
\end{equation} where $\omega'={\omega}\Big{(}1+{n}/{|\alpha(0)|^{2}}\Big{)}/2$.
This solution and all subsequent solutions are discussed in Sec.
\ref{4c}.

\subsubsection{When $\ket{\Psi(t)}$ is a coherent state}

If we assume that initially the system in is a coherent state
$\ket{\Psi(0)}=\ket{\beta}$, then by \erf{schExample},
\begin{equation} \label{coherentstate}
  \ket{\Psi(t)}=\exp(|\beta|^{2}/2)\sum_{n}\frac{ \beta^{n} e^{-i\omega
  n t}}{\sqrt{n!}}\ket{n}.
\end{equation}
 Substituting this into \erf{chi}, gives
\begin{eqnarray}
  \chi_{\psi}(x^{+}(t),y^{-}(t))&=&\beta e^{-i\omega
  t}\frac{\sum_{m}m
  (\alpha^{*}\beta e^{-i\omega
   t})^{m-1}/m!}
{\sum_{m} (\alpha^{*}\beta e^{-i\omega
  t})^{m}/m!}\nn \\ &=&\beta e^{-i\omega
  t}.
\end{eqnarray}
Thus
\begin{equation}
   d_t{\alpha}(t)=-\frac{i\omega}{2}\alpha(t)-\frac{i\omega}{2}\beta e^{-i\omega
  t}.
\end{equation} This has the solution
\begin{equation}\label{alphacoh}
  \alpha(t)=[\alpha(0)-\beta]e^{-i\omega t/2}+\beta e^{-i\omega
  t}.
\end{equation}

\subsection{Bohmian Mechanics, The Position Projector} \label{Bohmian}

In this section we consider the modal dynamics which describe a
decomposition into position eigenstates. That is the preferred
projective measure is $\{\hat{\pi}(x)dx=\ket{x}\bra{x}dx\}$.
 Since this is already a projector, there is no need to enlarge the
universe, and as shown above (and in Ref. \cite{Sud87} and
Ref. \cite{Vin93}) the modal dynamics for this case is just
Bohmian mechanics.

In terms of $\hat{X}$ and its conjugate operator $\hat{Y}$, the
Hamiltonian for the universe, \erf{HuniExample}, becomes
$\hat{H}_{\rm uni}(t)$
\begin{equation}\label{Hunipos}
  \hat{H}_{\rm uni}(t)=\omega (\hat{X}^{2}+\hat{Y}^{2}-1)/2.
\end{equation}
Using our velocity operator technique it can easily be shown that
the velocity field is
\begin{equation}\label{velocityfieldpos}
  v(x,t)=\frac{\omega{\rm Re}[-i
\langle\Psi(t)\ket{x}\partial_{x} \bra{x}\Psi(t)\rangle]}{
\langle\Psi(t)\ket{x}\bra{x}\Psi(t)\rangle},
\end{equation}
as $\hat{v}=\omega \hat{Y}$. Using \erf{diff} this gives
\begin{equation}\label{xdotpos}
  d_t {x}(t)=\frac{\omega{\rm Im}[
\langle\Psi(t)\ket{x}\partial_{x} \bra{x}\Psi(t)\rangle]}{
\langle\Psi(t)\ket{x}\bra{x}\Psi(t)\rangle}\Big{|}_{x=x(t)}.
\end{equation}

Since $\hat{Y}$ does not commute with $\hat{X}$, we can not give
both $\hat{X}$ and $\hat{Y}$ definite status (to give $\hat{Y}$
property status we would have to chose a momentum projective
measure as the preferred projective measure). However, as in
Bohmian mechanics, we can define a momentum field, $y(x,t)$, by
\begin{eqnarray}\label{momet}
  y(x,t)&=&\frac{{\rm
  Re}[\bra{\Psi(t)}x\rangle\bra{x}\hat{y}\ket{\Psi(t)}]}
  {\bra{\Psi(t)}x\rangle\langle{x}\ket{\Psi(t)}} \nn \\ &=&\frac{{\rm Im}[
\langle\Psi(t)\ket{x}\partial_{x} \bra{x}\Psi(t)\rangle]}{
\langle\Psi(t)\ket{x}\bra{x}\Psi(t)\rangle}.
\end{eqnarray}
One interprets this momentum field as follows: if the system has
the position $x(t)$ then its momentum is $y(x)|_{x=x(t)}$. With
this and position $x(t)$ we can define a phase point ($x(t)$,
$y(x)|_{x=x(t)}$), which in complex notation is written as
\begin{equation}\label{alphax}
  \alpha(t)=\frac{x(t)+iy(x)|_{x=x(t)}}{\sqrt{2}}.
\end{equation}The factor $1/\sqrt{2}$ is to
scale this to agree with the preceding section.

\subsubsection{When $\ket{\Psi(t)}$ is a number state}

As before when we assume a number state initial condition the
guiding wave at time $t$ is $\ket{\Psi(t)}=e^{-i\omega n
t}\ket{n}$. Substituting this into \erf{xdotpos} and using
\begin{equation}\label{xstate}
  \ket{x}=\frac{1}{\pi^{1/4}}\exp(-x^{2}/2)\sum_{n}\frac{H_n(x)}{\sqrt{2^n n!}}\ket{n}
\end{equation} where $H_n(x)$ is a $n^{\rm th}$ order Hermite
polynomial,
\begin{eqnarray}\label{xdotposnum}
  d_t {x}(t)&=&\frac{\omega{\rm Im}\{
\partial_{x}[\exp(-x^{2}/2
) H_{n}(x)]/\sqrt{2^n n!}\}}{\exp(-x^{2}/2 ) H_{n}(x)/\sqrt{2^n
n!}}\Big{|}_{x=x(t)}\nn \\ &=&0.
\end{eqnarray}
Using a similar argument and \erf{momet} we get $
y(x,t)_{x=x(t)}=0.$ That is, once the initial $x(0)$ is picked
from the quantum mechanical distribution is stays there for all
time. In terms of the complex notation $\alpha(t)$ we get
\begin{equation}\label{alphanumx}
  {\alpha}(t)=x(0)/\sqrt{2}.
\end{equation}

\subsubsection{When $\ket{\Psi(t)}$ is a coherent state}

If we assume that initially the system in is a coherent state
$\ket{\Psi(0)}=\ket{\beta}$, then \erf{coherentstate} in the
position representation is
\begin{eqnarray}
  \ket{\Psi(t)}&=&\frac{\exp(|\beta|^{2}/2)}{\pi^{1/4}}\int dx'
  \exp(\sqrt{2}\beta e^{-i\omega t}x\nl-\beta^{2} e^{-i2\omega
  t}/2)\exp(-x'^{2}/2)\ket{x'}.
\end{eqnarray}
Substituting this into \erf{xdotpos} gives
\begin{eqnarray}\label{xdotposcoh}
  d_t {x}(t)&=&\frac{\omega{\rm Im}\{
\partial_{x}[\exp(-x'^{2}/2)\exp(\sqrt{2}\beta e^{-i\omega t}x)]\}}{\exp(-x'^{2}/2)\exp(\sqrt{2}\beta e^{-i\omega
t}x) }\Big{|}_{x=x(t)}\nn\\&=&\omega\sqrt{2}{\rm Im}[\beta
e^{-i\omega t}],
\end{eqnarray}
and using \erf{momet} we get $  y(x,t)_{x=x(t)}=\sqrt{2}{\rm
Im}[\beta e^{-i\omega t}].$ Taking the derivative of this gives
\begin{equation}
   d_t{y}(x,t)_{x=x(t)}=-\omega\sqrt{2}{\rm Re}[\beta e^{-i\omega
  t}].
\end{equation} Thus
\begin{equation}
   d_t{\alpha}(t)=\omega\{{\rm Im}[\beta
e^{-i\omega t}]-i{\rm Re}[\beta e^{-i\omega t}]\}=-i\omega\beta
e^{-i\omega t}.
\end{equation} This has the solution
\begin{equation}\label{alphacohx}
  \alpha(t)=\beta e^{-i\omega t} +(\alpha(0)-\beta).
\end{equation}

\subsection{Classical Limit}\label{4c}

The classical harmonic oscillator has the well known solution,
this being
\begin{equation}\label{alphac}
  {\alpha}(t)=\alpha(0)e^{-i\omega t}.
\end{equation}
In Sec.~\ref{Husimi} (modal dynamics when the preferred POM is
the Husimi POM) we saw that in the enlarged Hilbert space, we can
define $\alpha(t)=x^{+}(t)+iy^{-}(t)$. Now consider the classical
limit. First  consider the case of the number state
\erf{alphanum}.
 From the probability formula (\ref{ProbPOM3}) it can be shown that
 $|\alpha(t)|^{2} \approx n$ with high probability.
This means that as $n\to \infty$, $\omega' \to \omega$, reproducing the classical dynamics.
  When considering the case with
an initial coherent state we can similarly argue that $\beta
\approx
 \alpha(0)$ with high probability.  Then in the limit $|\beta| \to \infty$, the difference between
the classical formula and  \erf{alphacoh} is negligible.

By contrast, in the position case (Bohmian mechanics), for the
first case (number state), ${\alpha}(t)=x(0)$ for all $n$, the
dynamics is completely non-classical. However, it can be argued
that when we consider the second case with an initial coherent
state, again the difference between the classical formula and
\erf{alphacohx} is negligible. The fact that in both Bohmian
mechanics and the modal interpretation of Sec.~\ref{Husimi} have a
good classical correspondence for coherent state is not
surprising, as the coherent state is a classical-like state. What
is surprising is that it is possible to obtain classical modal
dynamics even for a non-classical state, by using POMs.

\section{DISCUSSION} \label{Sec5}

In this paper we have extended the modal dynamics of Refs.~
\cite{Bel84,Sud87,Vin93,Sud00,Dic97,BacDic99} to include the
possibility of having a preferred POM. To do this we enlarged the
Hilbert space from ${\cal H}_{\rm uni}$ to  ${\cal K}={\cal
H}_{\rm uni}\otimes{\cal H}_{\rm aux}$. Once in this enlarged
Hilbert space we used Naimark's theorem \cite{Hel76,BusGraLah95}
to define a preferred projective measure which is equivalent to
the preferred POM. That is the statistics of an observable
describe by ${Z}=\{(z_{n},\hat{F}_{n}(t))\}$, (where
$\hat{F}_{n}(t)$ is a POM element) is equivalent to the statistic
of the property ${Z}=\{(z_{n},\hat{\Pi}_{n}(t))\}$. Here
  $\hat{\Pi}(t)$ is a projector in ${\cal K}$ and defines the
property state $\ket{\Phi_{n}(t)}$. This state represents the
actual state of the universe and determines the present value,
$z_n$, of the property $\hat{Z}$ from the set of possible values.
The evolution (jumping between $z_n$) is determined by the
stochastic evolution of $\ket{\Phi_{n}(t)}$, which in-turn depends
on the guiding wave $\ket{\Phi(t)}=\ket{\phi}\ket{\Psi(t)}$.
$\ket{\Psi(t)}$ is the standard quantum state found from the \sch
equation and $\ket{\phi}$ is a state defined in the auxiliary
Hilbert space.

To illustrate modal dynamics for POMs we considered a simple
example: a universe consisting of a single Harmonic oscillator. To
illustrate our new dynamics, we looked at the Husimi POM and
compared the dynamics obtained to that which is obtained with the
position projective measure (Bohmian mechanics). Since the first
case corresponds to a POM we have to use our enlarged Hilbert
space modal dynamics to develop the stochastic evolution equation
for the value of the property (or equivalently the property
state). For the Husimi POM we denoted this value by $\alpha(t)$.
We find that by choosing a Naimark projector $\ket{\phi}=\ket{0}$,
this property defines a point in phase space
($\alpha(t)=x^{+}+iy^{-}$) that is effectively the position and
momentum of the system.
  Investigating two different initial conditions for $\ket{\Psi(0)}$, namely
a number state and coherent state, we find that the differential
equation for $\alpha(t)$ for both cases has a classical limit
which agrees with classical theories. When comparing to the modal
dynamics for
  the  position
projective measure we find that when the initial state is the
number state the dynamics are highly non-classical. Only for an
initial coherent state (a classical-like state) can
 a classical limit can be obtained.

In conclusion by extending modal dynamics to include POMs allows
the ability to include overcomplete decomposition, like the Husimi
POM \cite{Husimi,WalMil94} and  informationally complete POMs
\cite{Info,Fuchs}. This may provide an answer to questions
involving the quantum-classical limit. An interesting question for
future work is whether the extension of the Hilbert space is only
a mathematical tool or whether there is some physical significance
behind enlarging the Hilbert space.
 Apart
from this fundamental question we intend to use this theory to
explain diffusive non-Markovian stochastic schrodinger equations
(SSEs) \cite{DioGisStr98}. In a recent paper  we have shown that
under the orthodox interpretation non-Markovian SSEs represent
nothing more then a stencil which determines the state of a system
at a particular time $t$, given that a measurement on the
environment at that time yielded result $z$ \cite{GamWis02}. That
is, non-Markovian SSEs can not be interpreted as evolution
equations for the state of the system conditioned on the outcomes
of some continuous measurement of the environment. We believe that
under the modal theory it can be shown that non-Markovian SSEs are
evolution equations for the system part of the property state when
the environment property $Z$ is given definite status. In
\cite{GamWis03} we have shown that when the preferred measure is
position, non-Markovian SSEs do have this interpretation. To
interpret other non-Markovian SSEs \cite{DioGisStr98,GamWis02}, it
 is necessary to consider non-orthogonal decompositions.

\appendix

\section{Proof of \erf{NiaExample1}.}\label{A}

\begin{widetext}
To show that 
$\ket{\alpha}\bra{\alpha}/{\pi}=\langle{0}\ket{x^{+},y^{-}}\bra{x^{+},y^{-}}0\rangle$
we consider just
$\langle{0}\ket{x^{+},y^{-}}=\ket{\alpha}/\sqrt{\pi}$. Using
\erf{x+y-state}, and the standard definition of a $x$-state
\erf{xstate}, $\langle{0}\ket{x^{+},y^{-}}$ becomes
\begin{equation}
\langle{0}\ket{x^{+},y^{-}} =\int
\frac{dx'}{\sqrt{2\pi}}\frac{1}{\pi^{1/4}}\exp\Big{[}\frac{-(x^{+}-x')^{2}}{4}\Big{]}\Ket{\frac{x^{+}+x'}{\sqrt{2}}}_{\rm
uni}e^{iy^{-}x'}.
\end{equation} 
Defining $X=(x^{+}+x')/{\sqrt{2}}$ allows us to
rewrite this as
\begin{equation}
\langle{0}\ket{x^{+},y^{-}} =\int
\frac{dX}{\sqrt{\pi}}\frac{1}{\pi^{1/4}}\exp\Big{[}\frac{-(2x^{+}-\sqrt{2}X)^{2}}{4}\Big{]}
\ket{X}_{\rm uni}{\sqrt{2}}e^{iy^{-}(\sqrt{2}X-x^{+})},
\end{equation} which with definition \erf{xstate} can be expanded
to
\begin{eqnarray}
\langle{0}\ket{x^{+},y^{-}}& =&\sum_{n}\int \frac{dX}{{\pi}}\exp(
-x^{+}x^{+}+\sqrt{2}Xx^{+}-X^{2}+ iy^{-}\sqrt{2}X-iy^{-}x^{+})
\frac{H_n(X)}{\sqrt{2^n n!}}\ket{n}_{\rm uni}
\\ &=&\exp(-|\alpha|^{2}/2)\sum_{n}\int
\frac{dX}{\pi}\exp(-X^{2})\exp(2\alpha X/\sqrt{2}-\alpha^{2}/2)
\frac{H_n(X)}{\sqrt{2^n n!}}\ket{n}_{\rm uni}
\end{eqnarray} where $\alpha=x^{+}+iy^{-}$. Then using $\sum_{m} t^{m}H_m(x)/m!=\exp(2tx-t^2)$ this can be
written as
\begin{eqnarray}
\langle{0}\ket{x^{+},y^{-}}&=&\frac{\exp(-|\alpha|^{2}/2)}{\pi}
\sum_{n,m}\frac{\alpha^{m}}{\sqrt{2^m} m!\sqrt{2^n
n!}}\ket{n}_{\rm uni} \int {dX}\exp(-X^{2}){H_m(X)}{H_n(X)}
\end{eqnarray} and since $\int
{dX}\exp(-X^{2}){H_m(X)}{H_n(X)}=2^{n}n!\sqrt{\pi}\delta_{nm}$
this becomes
\begin{eqnarray}
\langle{0}\ket{x^{+},y^{-}}&=&\frac{\exp(-|\alpha|^{2}/2)}{\sqrt{\pi}}
\sum_{n}\frac{\alpha^{n}}{\sqrt{n!}}\ket{n}_{\rm
uni}=\frac{1}{\sqrt{\pi}}\ket{\alpha}_{\rm uni}.
\end{eqnarray}
Thus
$\ket{\alpha}\bra{\alpha}/{\pi}=\langle{0}\ket{x^{+},y^{-}}\bra{x^{+},y^{-}}0\rangle$.
\end{widetext}
\acknowledgments

We would like to thank R. Spekkens for useful discussions. We
would also like to thank an unnamed referee who provided very
useful comments. We believe these comments have improved the
readability of the paper. This work was supported by the ARC.


\end{document}